\documentclass[conference]{IEEEtran} 
\IEEEoverridecommandlockouts
\usepackage{cite}
\usepackage{amsmath,amssymb,amsfonts}
\usepackage{algorithmic}
\usepackage{graphicx}
\usepackage{textcomp}
\usepackage{xcolor}
\def\BibTeX{{\rm B\kern-.05em{\sc i\kern-.025em b}\kern-.08em
    T\kern-.1667em\lower.7ex\hbox{E}\kern-.125emX}}

\usepackage{BOONDOX-ds}

\usepackage{bbm}

\begin{document}

\title{Mathematics of Digital Hyperspace
\thanks{This material is based upon work supported by the Assistant Secretary of Defense for Research and Engineering under Air Force Contract No. FA8702-15-D-0001, National Science Foundation CCF-1533644, and United States Air Force Research Laboratory Cooperative Agreement Number FA8750-19-2-1000. Any opinions, findings, conclusions or recommendations expressed in this material are those of the author(s) and do not necessarily reflect the views of the Assistant Secretary of Defense for Research and Engineering, the National Science Foundation, or the United States Air Force. The U.S. Government is authorized to reproduce and distribute reprints for Government purposes notwithstanding any copyright notation herein.}
}

\author{\IEEEauthorblockN{Jeremy Kepner$^{1,2,3}$, Timothy Davis$^4$, Vijay Gadepally$^{1,2}$, Hayden Jananthan$^{1,5}$, Lauren Milechin$^{6}$
\\
\IEEEauthorblockA{$^1$MIT Lincoln Laboratory Supercomputing Center,  $^2$MIT Computer Science \& AI Laboratory, \\$^3$MIT Mathematics Department, $^4$Texas A\&M, $^5$Vanderbilt, $^6$MIT Dept. of Earth, Atmospheric, \& Planetary Sciences
}}}
\maketitle

\begin{abstract}
Social media, e-commerce, streaming video, e-mail, cloud documents, web pages, traffic flows, and network packets fill vast digital lakes, rivers, and oceans that we each navigate daily.  This digital hyperspace is an amorphous flow of data supported by continuous streams that stretch standard concepts of type and dimension.  The unstructured data of digital hyperspace can be elegantly represented, traversed, and transformed via the mathematics of hypergraphs, hypersparse matrices, and associative array algebra.  This paper explores a novel mathematical concept, the semilink, that combines pairs of semirings to provide the essential operations for graph analytics, database operations, and machine learning.  The GraphBLAS standard currently supports hypergraphs, hypersparse matrices, the mathematics required for semilinks, and seamlessly performs graph, network, and matrix operations.  With the addition of key based indices (such as pointers to strings) and semilinks, GraphBLAS can become a richer associative array algebra and be a plug-in replacement for spreadsheets, database tables, and data centric operating systems, enhancing the navigation of unstructured data found in digital hyperspace.
\end{abstract}

\begin{IEEEkeywords}
graphs, hypergraphs, hypersparse, networks, polystore, databases, algebra
\end{IEEEkeywords}

\section{Introduction}
Global usage of the Internet is expected to exceed 5 billion people\cite{Cisco2018-2023}.  The volume, velocity, and variety of Internet data continues to expand.  Social media, e-commerce, streaming video, e-mail, cloud documents, web pages, traffic flows, and network packets fill vast digital lakes, rivers, and oceans that we each navigate daily \cite{sawadogo2020data}.   Some of the most common manifestations of these data are in the form of spreadsheets, database tables, matrices, graphs, and networks.  The resulting digital hyperspace is an amorphous flow of data supported by continuous streams of these objects that stretch standard concepts of type and dimension.

  Fortunately, the unstructured data of digital hyperspace can be elegantly represented, traversed, and transformed via the mathematics of hypergraphs \cite{ghoshal2009random,mordeson2012fuzzy,shun2020practical}, hypersparse matrices \cite{buluc2008representation,kepner2011graph,kepner202075}, and associative array algebra \cite{Kepner2011p-ch1,KepnerGadepally2014,kepner2016associative,kepner2018mathematics}.  These mathematics have been implemented in a variety of software libraries, including the GraphBLAS standard \cite{Kepner2016graphblas,bulucc2017design,kumar2018graphblas,mattson2019lagraph} implemented in the C/Matlab/Octave/Python/Julia languages \cite{davis2018graph,chamberlin2018pygb,moreira2018implementing,davis2019algorithm} and the RedisGraph database \cite{cailliau2019redisgraph}; the C-MPI CombBLAS parallel library \cite{BulucGilbert2011}; and the D4M associative array library in Matlab/Octave/Python/Julia languages \cite{kepner2011massive,Kepner2012-ch1,chen2016julia,milechin2017d4m,milechin2018database} with database bindings to SciDB, Accumulo, and PostGreSQL \cite{Kepner2013,Kepner2014a,GadepallyEtAl2015,Hutchison2015,samsi2016benchmarking}.  The GraphBLAS standard has further enabled hardware acceleration of these mathematics via multithreading \cite{aznaveh2020parallel}, GPUs \cite{wang2019accelerating}, and special purpose accelerators \cite{song2010,song2014processor,song2016novel,jia2019dissecting,james2020ispd}.

  Linearity is a key property of these mathematics utilized by the above implementations to leverage extensive linear systems theory \cite{kepner2018mathematics}. From a performance perspective, linearity is often manifest through the distributive property
$$
  a \otimes (b \oplus c) = (a \otimes b) \oplus (a \otimes c)
$$
enabling the reordering of operations critical for effective parallel computation and distributed database query planning.  From a data perspective, linearity provides the additive identity and multiplicative annihilator
$$
  a \oplus \emph{0} = a ~~~~~~~~~~~~ a \otimes \emph{0} = \emph{0}
$$
eliminating the need to store \emph{0} entries (an essential property for efficient sparse computations).  If fact, in this context, the above properties can be used to define \emph{0} for the relevant value set, $V$, which may, or may not, be the standard arithmetic 0.

Collectively, these mathematical properties are defined by mathematical {\it semirings} that are directly supported by the aforementioned technologies.   The increasing use of semirings for the manipulation of digital data has led to frequent coupling of distinct semirings in graph analysis \cite{kepner2011graph}, databases \cite{kepner2016associative}, and machine learning computations \cite{kepner2017enabling,kumar2018ibm,davis2019write}.  This paper explores some of mathematical properties of coupled semirings, here referred to as {\it semilinks}, and offers up some potential paths forward to formalizing and applying this novel mathematics as natural extensions to existing technologies, such as the GraphBLAS standard.

The rest of this paper is organized as follows.  First, some mathematical preliminaries regarding hypergraphs, hypersparse matrices, and semirings are provided.  Associative arrays are then summarized.  Next, some general properties of semilinks are explored and some specific possible semilinks are investigated in the context of graph analysis, databases, and machine learning.  Finally, some recommendations and conclusions are provided.

\section{Mathematical Preliminaries}

\begin{figure}
\center{\includegraphics[width=1.0\columnwidth]{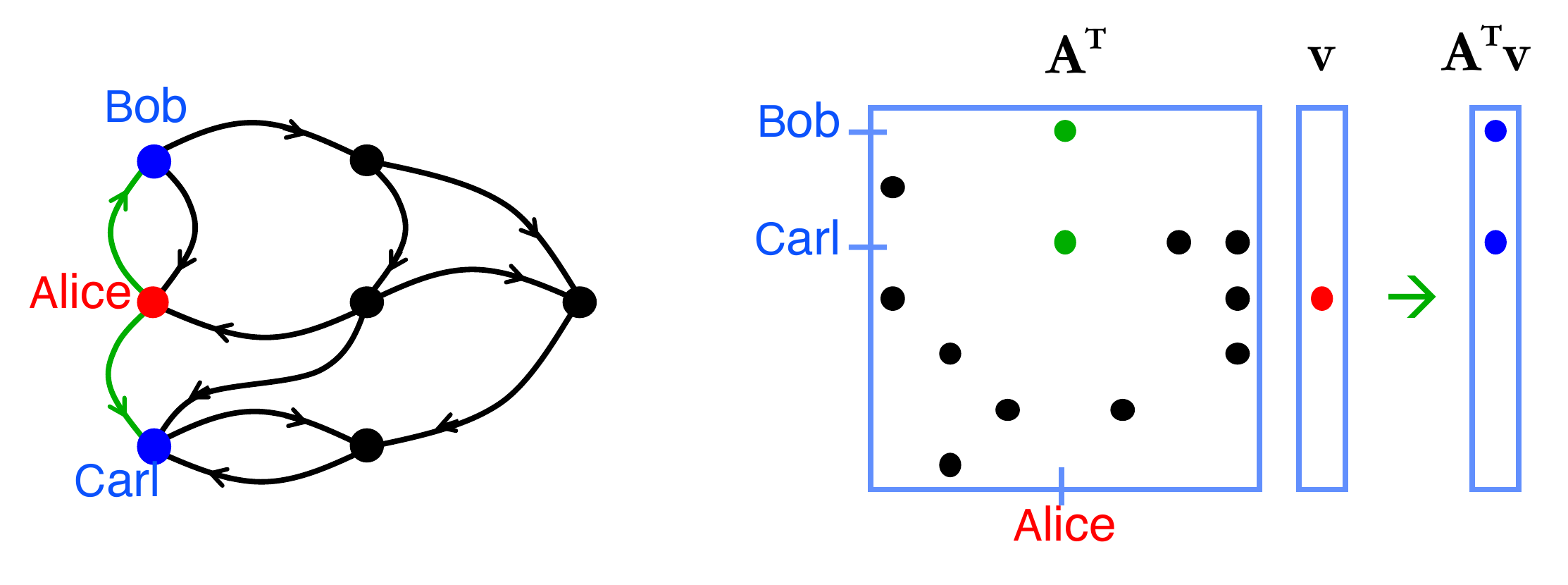}}
      	\caption{{\bf Graph Adjacency Array Duality}. Breadth-first-search performed on a graph (left) and an adjacency array (right) illustrates the deep connection between graphs and arrays.}
      	\label{fig:Adjacency}
\end{figure}

  The navigation of diverse digital data can be enhanced by a number of mathematical concepts which underpin the broader algebra of associative arrays which are briefly described in this section (see \cite{kepner2011graph} for a complete description).  Perhaps the most important is the graph-matrix duality illustrated in Figure~\ref{fig:Adjacency} that links the fundamental operation of graphs (breadth-first-search) with the fundamental operation of arrays (array multiplication), where an adjacency array
$$
  \mathbf{A}(k_1,k_2) \neq \emph{0}
$$
implies an edge from vertex $k_1$ to $k_2$.   Hypergraphs extend graphs to provide a natural representation of events that connect diverse entities.  Hypersparse arrays extend arrays to allow the efficient storage and operation on data that is growing without bounds.  Semirings extend standard arithmetic enabling operations on diverse data to utilize the power of linear systems theory.

\subsection{Hypergraphs}

\begin{figure}
\center{\includegraphics[width=1.0\columnwidth]{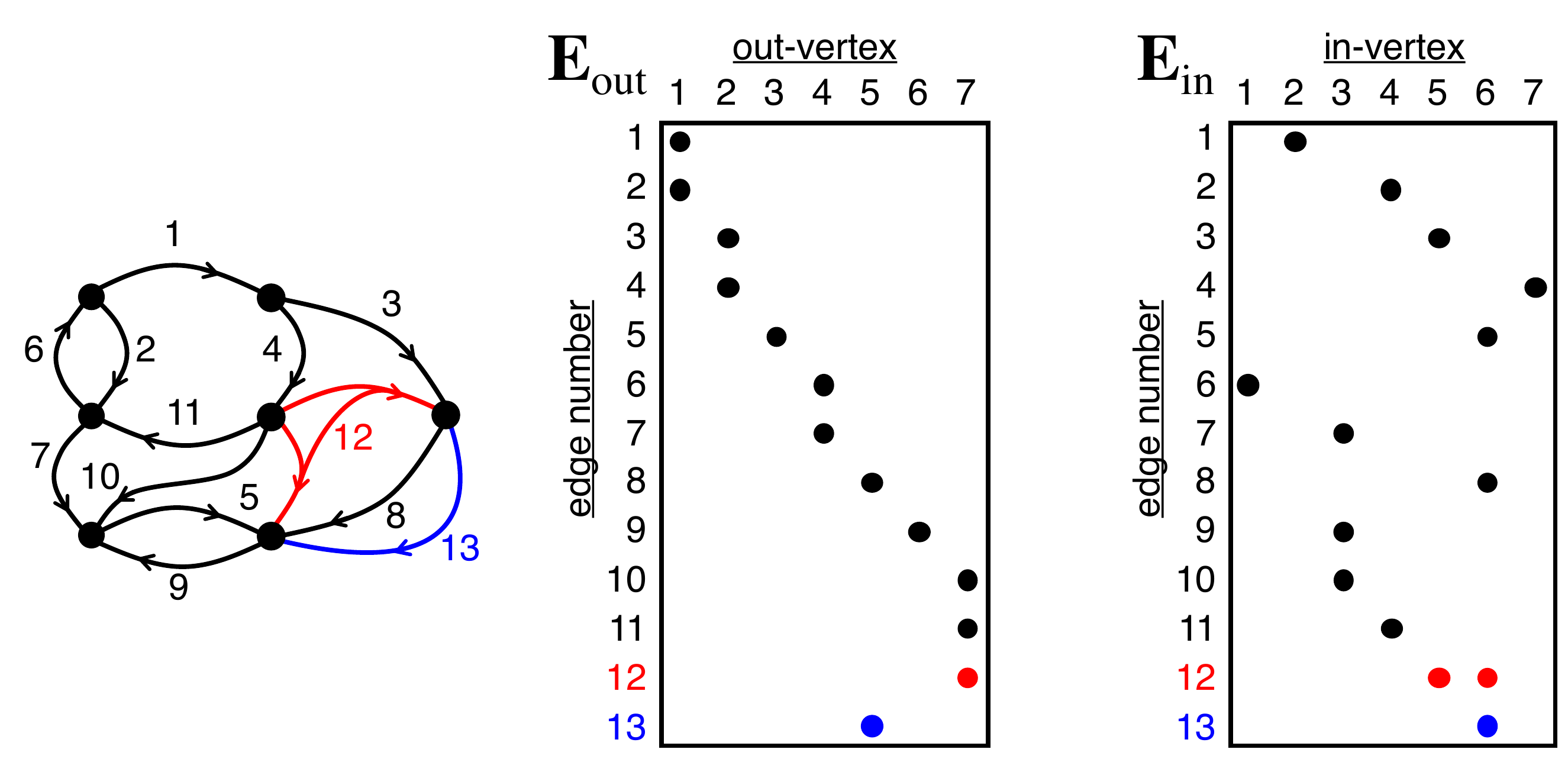}}
      	\caption{{\bf Hyper-Multi-Graph Edge Array Duality}. Incidence (or edge) arrays can capture hyper-edges (red) connecting more than two vertices and multi-edges (blue) between the same vertices.}
      	\label{fig:Edge}
\end{figure}

Adjacency arrays are a powerful tool for analyzing directed-weighted-graphs, but are unable to represent the diverse data that is  commonly found in streaming events.  These streaming events can be described as hyper-multi-weighted-directed-graphs and are best represented as incidence (or edge) arrays (see Figure~\ref{fig:Edge}), where 
$$
  \mathbf{E}_{\rm out}(k,k_1) \neq \emph{0} ~~~~~~ \mathbf{E}_{\rm in}(k,k_2) \neq \emph{0}
$$
implies that edge $k$ comes out of vertex $k_1$ and goes into vertex $k_2$.

\begin{figure}
\center{\includegraphics[width=1.0\columnwidth]{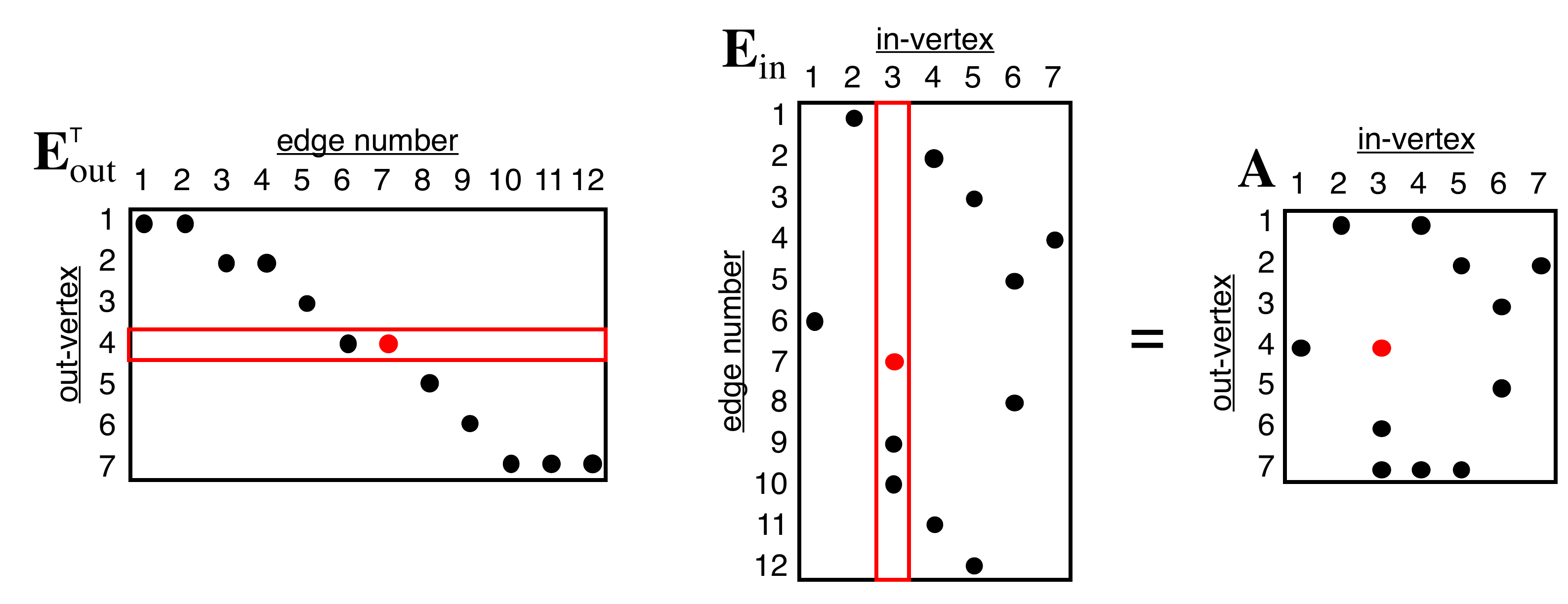}}
      	\caption{{\bf Edge Array to Adjacency Array}. Construction of an adjacency array of a graph from its incidence arrays via array multiply.  The entry $\mathbf{A}(4,3)$ is obtained by combining the row  $\mathbf{E}_\mathrm{out}^{\sf T}(4,k)$ with the column  $\mathbf{E}_\mathrm{in}(k,3)$ via the array product 
$
  \mathbf{A}(4,3)= \bigoplus\limits_{k = 1}^{12} \mathbf{E}_\mathrm{out}^{\sf T}(4,k) \otimes \mathbf{E}_\mathrm{in}(k,3)
$}
      	\label{fig:EdgeAdjacency}
\end{figure}

The adjacency array and the edge array are strongly coupled via array multiplication (Figure~\ref{fig:EdgeAdjacency})
$$
  \mathbf{A} = \mathbf{E}_\mathrm{out}^{\sf T} \mathbf{E}_\mathrm{in}
$$
where the individual values in $\mathbf{A}$ are computed via
$$
  \mathbf{A}(i,j)= \bigoplus\limits_{k} \mathbf{E}_\mathrm{out}^{\sf T}(i,k) \otimes \mathbf{E}_\mathrm{in}(k,j)
$$
The adjacency array represents a projection of edge data and is often an initial step in processing diverse digital data.

\subsection{Hypersparse}

\begin{figure}
\center{\includegraphics[width=1.0\columnwidth]{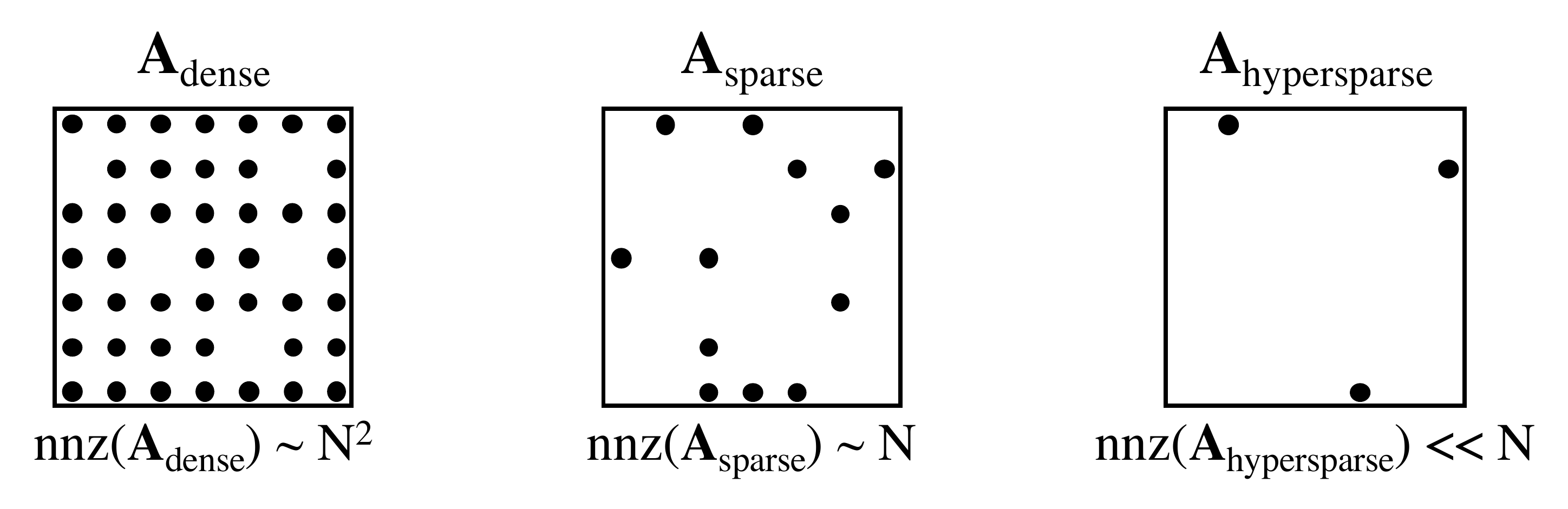}}
      	\caption{{\bf Dense, Sparse, and Hypersparse Arrays}. Sparsity concepts for an $N{\times}N$ array $\mathbf{A}$.}
      	\label{fig:Hypersparse}
\end{figure}

As the dimensions of digital data expand the concept of sparsity plays an increasing role.  Sensor data, such as images, are well presented by dense arrays where the number on non-zero entries $nnz()$ is small.  Physical networks, neural networks, mesh geometries, and other systems where the dimension of the problem is known can often be well represented by sparse arrays where $nnz()$ is on the order of the number of rows or columns in the array.  Data with dimensions that are continuously increasing can be captured by hypersparse arrays where $nnz()$ is much smaller than the number of rows or columns (Figure~\ref{fig:Hypersparse}).

\subsection{Semirings}

Obtaining the advantages of linear systems on diverse data involves extending  addition $\oplus$ and multiplication $\otimes$ beyond standard real numbers to include sets and strings.  If the set of values is denoted by $V$, then pairs of operations $\otimes$ and $\oplus$ that obey the distributive property on values from $V$ will generally exhibit the desired properties of a linear system.  Formally, the mathematical object with the desired mathematical properties is a semiring denoted $(V,\oplus,\otimes,\emph{0},\emph{1})$, where $\emph{0}$ is the $\oplus$ identity and $\emph{1}$ is the $\otimes$ identity.  Some of the common combinations of addition and multiplication operations that have proven valuable are standard arithmetic addition and multiplication ${+}.{\times}$, union and intersection ${\cup}.{\cap}$ in relational databases \cite{codd1970relational,maier1983theory,abiteboul1995foundations}, and various tropical algebras that are important in finance \cite{klemperer2010product,baldwin2016understanding} and neural networks \cite{kepner2017enabling,kumar2018ibm,davis2019write}: ${\max}.{+}$, ${\min}.{+}$, ${\max}.{\times}$, ${\min}.{\times}$, ${\max}.{\min}$, and ${\min}.{\max}$.   Examples of commonly used semirings are shown in Table~\ref{tab:Semirings}.  For a guide to the literature on semirings and their applications see \cite{glazek2002guide}.

\begin{table}
\caption{Selected Semirings}
\vspace{-0.25cm}
Some semirings that play important roles in many real-world applications. $\mathbb{R}$ are the real numbers.  $\mathbb{R}_{\geq 0}$ are the non-negative real numbers. $\mathbb{V}$ is any strict totally ordered set (i.e., sortable).  $\mathcal{P}()$ is the power set (set of all subsets).  $\emptyset$ is the empty set.   +$\infty$ is the maximum element of a set.   -$\infty$ is the minimal element of a set.
\begin{center}
\begin{tabular}{lcccc}
\hline
Set & $\oplus$ & $\otimes$ & \emph{0} & \emph{1}  \\
\hline
$\mathbb{R}$ & $+$ & $\times$ & 0 & 1  \\
$\mathbb{R}~\cup$ -$\infty$ & $\max$ & $+$ & -$\infty$ & 0  \\
$\mathbb{R}~\cup$ +$\infty$ & $\min$ & $+$ & +$\infty$ & 0  \\
$\mathbb{R}_{\geq 0}$ & $\max$ & $\times$ & 0 & 1  \\
$\mathbb{R}_{\geq 0}~\cup$ +$\infty$ & $\min$ & $\times$ & +$\infty$ & 1  \\
$\mathbb{V}$ & $\cup$ & $\cap$ & $\emptyset$ & $\mathcal{P}(\mathbb{V})$  \\
$\mathbb{V}~\cup$ -$\infty$ & $\max$ & $\min$ & -$\infty$ & +$\infty$  \\
$\mathbb{V}~\cup$ +$\infty$ & $\min$ & $\max$ & +$\infty$ & -$\infty$  \\
\hline
\end{tabular}
\end{center}
\label{tab:Semirings}
\end{table}%

\section{Associative Arrays}

  The full mathematics of associative arrays and the ways they build on the mathematics of the previous section to encompass spreadsheets, database tables, matrices, graphs, networks, and higher dimension tensors are fully described in \cite{Kepner2011p-ch1,KepnerGadepally2014,kepner2016associative,kepner2018mathematics}.  Only the essential mathematical properties of associative arrays are reviewed here.  The essence of associative array algebra is three operations: element-wise addition $\oplus$, element-wise multiplication $\otimes$, and array multiplication ${\oplus}.{\otimes}$.  In brief, the set of associative arrays are defined as a mapping from sets of keys to values
$$
  \mathbb{A}: K_1 \times K_2 \to \mathbb{V}
$$
where $K_1$ (the set of row keys)  and $K_2$ (the set of column keys) can be any sortable sets, such as the integers, real numbers, or strings.  $\mathbb{V}$ is a set of values that forms a semiring ($\mathbb{V}$,$\oplus$,$\otimes$,{\emph 0},{\emph 1}) with addition operation $\oplus$, multiplication operation $\otimes$, additive identity/multiplicative annihilator {\emph 0}, and multiplicative identity {\emph 1}. The values can take on many forms, such as numbers, strings, and sets.

Associative array algebra and its specialization in the GraphBLAS reference two main semirings.  The first is the element-wise commutative semiring $(\mathbb{A},\oplus,\otimes,\mathbs{0},\mathbs{1})$ built from the two commutative monoids
$$
    M_\mathbs{0} = (\mathbb{A},\oplus,\mathbs{0})
    ~~~~~~~~~~
    M_\mathbs{1} = (\mathbb{A},\otimes,\mathbs{1})
$$
where $\mathbs{0}$ is the array of all \emph{0} and  $\mathbs{1}$ is the array of all \emph{1}.  Likewise, the array semiring $(\mathbb{A},\oplus,{\oplus}.{\otimes},\mathbs{0},\mathbb{I})$ built from a commutative monoid and non-commutative monoid
$$
    M_\mathbs{0} = (\mathbb{A},\oplus,\mathbs{0})
    ~~~~~~~~~~
    M_\mathbb{I} = (\mathbb{A},{\oplus}.{\otimes},\mathbb{I})
$$
where $\mathbb{I}(k,k) = \emph{1}$ and $\emph{0}$ otherwise.  Many of the properties of associative arrays that will be utilized in the semilink discussion are listed in Table~\ref{tab:AssociativeArray}.  Of particular practical importance are the large row and column key spaces typically used in associative arrays that practically eliminate the dimensional conformance rules required in matrix operations.  As a result, associative arrays are typically added and multiplied with little regard for the true dimensions of their large row and column key spaces.  What is more important to producing non-trivial results that are not all \emph{0} is some overlap in the non-zero row and column keys of the constituent associative arrays.

\begin{table}
\caption{Associative Arrays}
\vspace{-0.25cm}
Summary of associative array operations and properties. $\mathbf{k}$, $\mathbf{k}_1$, $\mathbf{k}_2$, and $\mathbf{v}$ are vectors of the row keys, column keys, and values of the nonzero elements of the associative array $\mathbf{A}$.  $\mathbs{0}$ is an array of all \emph{0}. $\mathbs{1}$ is an array of all \emph{1}.  $|~|_0$ is the element-wise zero-norm that maps all non-zero elements to \emph{1}.
\begin{center}
\begin{tabular}{p{0.75in}p{2.4in}}
\hline
{\bf Property} & {\bf Notation} \\
\hline
Construction & $\mathbf{A} = \mathbb{A}(\mathbf{k}_1,\mathbf{k}_2,\mathbf{v})$ \\
Extraction & $(\mathbf{k}_1,\mathbf{k}_2,\mathbf{v}) = \mathbf{A}$ \\
Permutation & $\mathbb{P}(\mathbf{k}_1,\mathbf{k}_2) = \mathbb{A}(\mathbf{k}_1,\mathbf{k}_2,1)$  ~~~~~~~~~~~ $\mathbf{k}_1$, $\mathbf{k}_2$ unique \\
Identity & $\mathbb{I}(\mathbf{k}) = \mathbb{P}(\mathbf{k},\mathbf{k})$  \\
Transpose & $\mathbf{A}(k_2,k_1) = \mathbf{A}^{\sf T}(k_1,k_2)$  \\
Row keys & $\mathbf{k}_1 = {\rm row}(\mathbf{A})$  ~~~~~~~~~~~~~~~~~~~~~~~~~~~~~~~~~ $\mathbf{k}_1$ unique\\
Column keys & $\mathbf{k}_2 = {\rm col}(\mathbf{A})$  ~~~~~~~~~~~~~~~~~~~~~~~~~~~~~~~~~ $\mathbf{k}_2$ unique\\
Nonzero count & ${\rm nnz}(\mathbf{A})$ \\
Same sparsity & $|\mathbf{A}|_0 = |\mathbf{B}|_0$  \\
\hline
Element-wise  & $\mathbf{C} = \mathbf{A} \oplus \mathbf{B}$~~~~~~~~~~~~~~~$\mathbf{A} \oplus \mathbs{0} = \mathbf{A}$\\
~ addition & $\mathbf{C}(k_1,k_2) = \mathbf{A}(k_1,k_2) \oplus \mathbf{B}(k_1,k_2)$  \\
\hline
Element-wise  & $\mathbf{C} = \mathbf{A} \otimes \mathbf{B}$~~~~~~~~~~~~~~~~$\mathbf{A} \otimes \mathbs{1} = \mathbf{A}$~~~~$\mathbf{A} \otimes \mathbs{0} = \mathbs{0}$\\
~ multiplication & $\mathbf{C}(k_1,k_2) = \mathbf{A}(k_1,k_2) \otimes \mathbf{B}(k_1,k_2)$  \\
\hline
Array  & $\mathbf{C} = \mathbf{A} \mathbf{B} = \mathbf{A} {\oplus}.{\otimes} \mathbf{B}$~~~~~~~~~$\mathbf{A} \mathbb{I} = \mathbf{A}$~~~~~~~~~$\mathbf{A} \mathbs{0} = \mathbs{0}$ \\
~ multiplication & $\mathbf{C}(k_1,k_2) = \bigoplus_k \mathbf{A}(k_1,k) \otimes \mathbf{B}(k,k_2)$  \\
\hline
Commutativity & $\mathbf{A} \oplus \mathbf{B} = \mathbf{B} \oplus \mathbf{A}$ \\
              & $\mathbf{A} \otimes \mathbf{B} = \mathbf{B} \otimes \mathbf{A}$ \\
              & $(\mathbf{A} \mathbf{B})^{\sf T} = \mathbf{B}^{\sf T} \mathbf{A}^{\sf T}$ \\
\hline
Associativity & $(\mathbf{A} \oplus \mathbf{B}) \oplus \mathbf{C} = \mathbf{A} \oplus (\mathbf{B} \oplus \mathbf{C})$ \\
              & $(\mathbf{A} \otimes \mathbf{B}) \otimes \mathbf{C} = \mathbf{A} \otimes (\mathbf{B} \otimes \mathbf{C})$ \\
              & $(\mathbf{A}  \mathbf{B})  \mathbf{C} = \mathbf{A}  (\mathbf{B}  \mathbf{C})$ \\
\hline
Distributivity & $\mathbf{A} \otimes (\mathbf{B} \oplus \mathbf{C}) = (\mathbf{A} \otimes \mathbf{B}) \oplus (\mathbf{A} \otimes \mathbf{C})$ \\
              & $\mathbf{A} (\mathbf{B} \oplus \mathbf{C}) = (\mathbf{A}  \mathbf{B}) \oplus (\mathbf{A}  \mathbf{C})$ \\
\hline
\end{tabular}
\end{center}
\label{tab:AssociativeArray}
\end{table}%

\section{Semirings to Semilinks}

The overlap between three monoids and two semirings commonly used in associative arrays suggests investigating them as a potentially new mathematical concept referred to here as a \emph{semilink}
\begin{equation*}
(\mathbb{A},\oplus,\otimes,{\oplus}.{\otimes},\mathbs{0},\mathbs{1},\mathbb{I})
\end{equation*}
Among the standard (albeit somewhat rare) algebraic structures admitting three binary operations are residuated lattices \cite{blount2003structure}, Poisson algebras \cite{aguiar2000pre}, exponential fields \cite{kuhlmann2000ordered}, and quasigroups \cite{smith2006introduction}.   The closest in flavor to our semilink that of a composition ring, though even when working over a ring or field the semilink above does not satisfy the identities required to be a composition ring. In addition to being closed under any combination of operations $\oplus$, $\otimes$, and ${\oplus}.{\otimes}$ on associative arrays, such a combination of monoids/semirings would seem to have several properties. As part of semirings the pairs of operations $(\oplus, \otimes)$  and $(\oplus, {\oplus}.{\otimes})$ retain their properties within each pair such as the distributive property and the additive identity is the multiplicative annihilator. Important questions with regards to a semilink are what properties might exist among the pair of operations $(\otimes, {\oplus}.{\otimes})$ and their respective identities $\mathbs{1}$ and $\mathbb{I}$.

It is readily observable that the identities $\mathbs{1}$ and $\mathbb{I}$ preserve their properties with respect to their corresponding operations.  For example
$$
   \mathbs{1} \otimes \mathbb{I} = \mathbb{I} \otimes \mathbs{1} = \mathbb{I}
   ~~~~~~~~~~~
   \mathbs{1} {\oplus}.{\otimes} \mathbb{I} = \mathbb{I} {\oplus}.{\otimes} \mathbs{1} = \mathbs{1}
$$
$\mathbb{I}$ behaves like an identity under $\otimes$ if the array matches the sparsity structure of $\mathbb{I}$.  If $|\mathbf{A}|_0 = \mathbb{I}$, then
$$
    \mathbf{A} \otimes \mathbb{I} = \mathbb{I} \otimes \mathbf{A} = \mathbf{A}
$$
where $|~|_0$ is the element-wise zero-norm that maps all non-zero elements to \emph{1}. More generally, if the sparsity pattern of $\mathbf{A}$ is a permutation $|\mathbf{A}|_0 = \mathbb{P}$, then
$$
    \mathbf{A} \otimes \mathbb{P} = \mathbb{P} \otimes \mathbf{A} = \mathbf{A}
$$
In contrast, $\mathbs{1}$ with ${\oplus}.{\otimes}$ projects an array onto its rows or columns
$$
    \mathbf{C} = \mathbf{A} {\oplus}.{\otimes} \mathbs{1} \implies \mathbf{C}(k_1,:) = \bigoplus_{k_2} \mathbf{A}(k_1,k_2)
$$
$$
    \mathbf{C} =   \mathbs{1} {\oplus}.{\otimes} \mathbf{A} \implies \mathbf{C}(:,k_2) = \bigoplus_{k_1} \mathbf{A}(k_1,k_2)
$$
Interestingly, under certain conditions, ${\oplus}.{\otimes}$ distributes over ${\otimes}$.  Specifically, if $\mathbf{A}$ has the sparsity pattern of a permutation
$$
  |\mathbf{A}|_0 = |\mathbf{A}_1|_0 = |\mathbf{A}_2|_0 = \mathbb{P}
$$
and $\mathbf{A} = \mathbf{A}_1 \otimes \mathbf{A}_2$, then
$$
    \mathbf{A} {\oplus}.{\otimes} (\mathbf{B} \otimes \mathbf{C}) =
      (\mathbf{A}_1 {\oplus}.{\otimes} \mathbf{B}) \otimes (\mathbf{A}_2 {\oplus}.{\otimes} \mathbf{C})
$$
Similarly, a hybrid associativity does hold in the trivial case.  If $\mathbf{A} = \mathbs{1}$  or $\mathbf{C} = \mathbb{I}$, then
$$
     \mathbf{A} \otimes (\mathbf{B} {\oplus}.{\otimes} \mathbf{C}) = (\mathbf{A} \otimes \mathbf{B}) {\oplus}.{\otimes} \mathbf{C}
$$
In a related result, if the non-zero entries of $\mathbf{A}$, $\mathbf{B}$, and $\mathbf{C}$ do not have sufficient overlap, then the result will be $\mathbs{0}$. Using the row() and col() functions defined in Table~\ref{tab:AssociativeArray}, if
\begin{eqnarray*}
    {\rm row}(\mathbf{A}) \cap {\rm row}(\mathbf{B}) &=& \emptyset ~~~~~~ {\rm or} \\
    {\rm col}(\mathbf{A}) \cap {\rm col}(\mathbf{C}) &=& \emptyset ~~~~~~ {\rm or} \\
    {\rm col}(\mathbf{B}) \cap {\rm row}(\mathbf{C}) &=& \emptyset 
\end{eqnarray*}
then
$$
    \mathbf{A} \otimes (\mathbf{B} {\oplus}.{\otimes} \mathbf{C}) = \mathbs{0}
$$
Likewise, if
\begin{eqnarray*}
    {\rm row}(\mathbf{A}) \cap {\rm row}(\mathbf{B}) &=& \emptyset ~~~~~~ {\rm or} \\
    {\rm col}(\mathbf{A}) \cap {\rm col}(\mathbf{B}) &=& \emptyset ~~~~~~ {\rm or} \\
    {\rm col}(\mathbf{A}) \cap {\rm row}(\mathbf{C}) &=& \emptyset ~~~~~~ {\rm or} \\
    {\rm col}(\mathbf{B}) \cap {\rm row}(\mathbf{C}) &=& \emptyset
\end{eqnarray*}
then
$$
    (\mathbf{A} \otimes \mathbf{B}) {\oplus}.{\otimes} \mathbf{C} = \mathbs{0}
$$
which implies that if
\begin{eqnarray*}
    {\rm row}(\mathbf{A}) \cap {\rm row}(\mathbf{B}) &=& \emptyset ~~~~~~ {\rm or} \\
    {\rm col}(\mathbf{B}) \cap {\rm row}(\mathbf{C}) &=& \emptyset
\end{eqnarray*}
then
$$
     \mathbf{A} \otimes (\mathbf{B} {\oplus}.{\otimes} \mathbf{C}) = (\mathbf{A} \otimes \mathbf{B}) {\oplus}.{\otimes} \mathbf{C} = \mathbs{0}
$$

\section{Examples}

An important motivation for exploring the semilink concept is their common occurrence in practical applications.     In this section several semilinks are explored in the context of graphs, databases, and neural networks.

\subsection{Graph Analytics}

\begin{figure}[htb]
  	\centering
    	\includegraphics[width=\columnwidth]{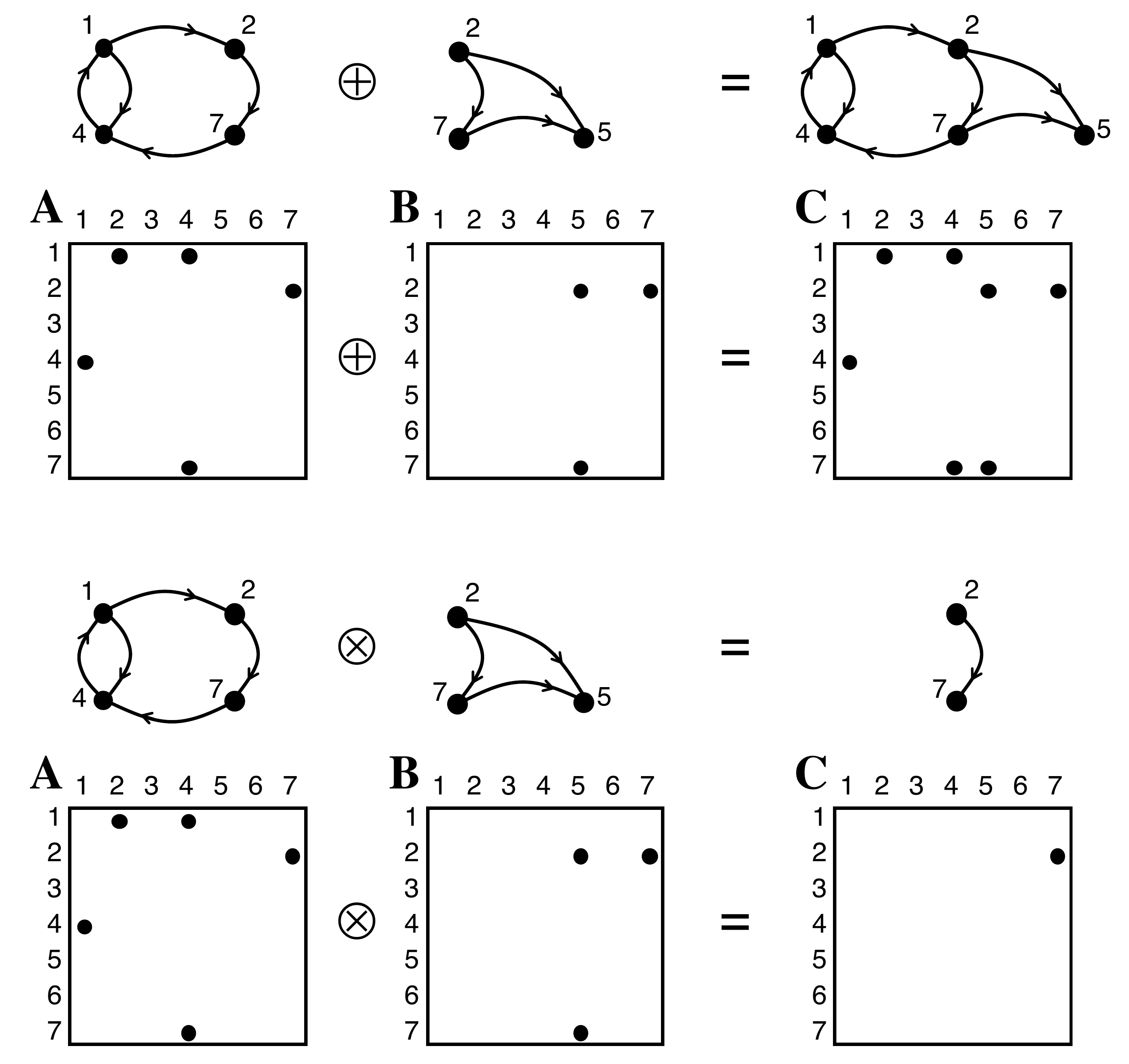}
      	\caption{{\bf Graph Union and Intersection}. (top) Element-wise addition $\oplus$ of associative arrays corresponds to graph union.  (bottom) Element-wise multiplication $\otimes$ of associative arrays corresponds to graph intersection.}
      	\label{fig:UnionIntersection}
\end{figure}

The general semilink
$$
    (\mathbb{A},\oplus,\otimes,{\oplus}.{\otimes},\mathbs{0},\mathbs{1},\mathbb{I})
$$
covers a number of important operations in graph analysis. Figure~\ref{fig:Adjacency} illustrates the duality between the fundamental operation of graphs (breadth-first-search) and the fundamental operation of arrays (array multiplication) ${\oplus}.{\otimes}$.  Figure~\ref{fig:UnionIntersection} shows how element-wise addition $\oplus$ and element-wise multiplication $\otimes$ correspond to graph union and graph intersection, which are also important graph operations.  In these graph operations, the essence of the calculation is topological and is determined by the presence of non-zero values in the result and not the exact value itself.  Thus, the core topological aspects of graph breadth-first-search, graph union, and graph intersection operations hold for any semiring on the values of the corresponding associative array, including all the semirings listed in Table~\ref{tab:Semirings}.

\subsection{Database Operations}

\begin{figure*}[t]
\centering
\includegraphics[width=6in]{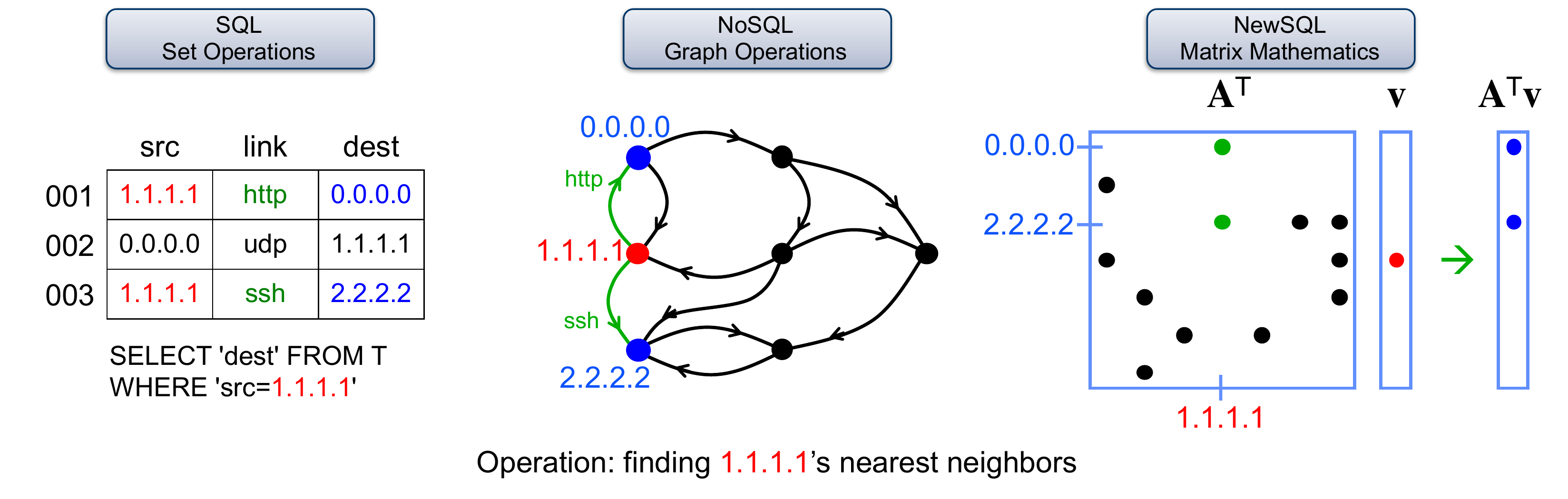}
\caption{Associative arrays combine the properties of databases, graphs, and matrices and provide common mathematics that span SQL, NoSQL, and NewSQL databases, and are ideal for analyzing networks.  The diagram shows the graph operation of finding the neighbors of 1.1.1.1 in each representation.}
\label{fig:AssociativeArrays}
\end{figure*}

Many database table operations can be mapped onto well-defined mathematical operations with known mathematical properties (see Figure~\ref{fig:AssociativeArrays}).  For example, relational (or SQL) databases \cite{Stonebraker1976,date1989guide,elmasri2010fundamentals} are described by relational algebra \cite{codd1970relational,maier1983theory,abiteboul1995foundations} that corresponds to the union-intersection semiring ${\cup}.{\cap}$ \cite{jananthan2017polystore}.  Triple-store databases (NoSQL) \cite{DeCandia2007,LakshmanMalik2010,George2011,7322476,Wall2015} and analytic databases (NewSQL) \cite{Stonebraker2005,Kallman2008,Balazinska2009,StonebrakerWeisberg2013,Hutchison2015,gadepally2015graphulo}  follow similar mathematics \cite{kepner2016associative}.  The table operations of these databases are further encompassed by associative array algebra, which brings the beneficial properties of array mathematics and sparse linear systems theory, such as closure, commutativity, associativity, and distributivity \cite{kepner2018mathematics}.  These mathematical properties provide strong correctness and linearity guarantees that are independent of scale and particularly helpful when trying to reason about massively parallel systems.

  The full mathematics of associative arrays and the ways they encompass  relational algebra are described in the aforementioned references \cite{kepner2016associative,jananthan2017polystore,kepner2018mathematics}. In brief, an associative array $\mathbf{A}$ is defined as a mapping from sets of keys to values.  The row keys are equivalent to the sequence ID in a relational database table.  The column keys are equivalent to the column names or record fields in a database table.  Intersection $\cap$ distributing over union $\cup$ is essential to database query planning and parallel query execution over partioned/sharded database tables \cite{booth1976distributed,shaw1980relational,stonebraker1986case,barroso2003web,curino2010schism,pavlo2012skew,corbett2013spanner}. 
  
  Perhaps the most canonical function in a relational database is the SQL {\sf select} statement that returns the columns $\mathbf{k}$ of rows in a table $\mathbf{A}$ that satisfy a specific condition, such as the value in column $\mathbf{k}(i)$ is $v$
$$
{\sf select}~\mathbf{k}(1),...,\mathbf{k}(n)~{\sf from}~\mathbf{A}~{\sf where}~\mathbf{k}(i) = v
$$
In terms of the associative array notation listed in Table~\ref{tab:AssociativeArray}, the above {\sf select} can be concisely written as
$$
  \mathbf{A}({\rm row}(\mathbf{A}(\mathbf{k}(i),:) = v),\mathbf{k})
$$
For many databases, the relevant semilink is
$$
    (\mathbb{A},\cup,\cap,{\cup}.{\cap},\emptyset,\mathbs{1},\mathbb{I})
$$
where each entry in $\mathbs{1}$ is $\mathcal{P}(\mathbb{V})$ and $\mathbb{I}(k,k) = \mathcal{P}(\mathbb{V})$ and $\emptyset$ otherwise.   The associative array version of the {\sf select} statement can be written in terms of  this semilink as
$$
  |((\mathbf{A} ~ {\cup}.{\cap} ~ \mathbb{I}(\mathbf{k}(i)) \cap v) ~ {\cup}.{\cap} ~ \mathbs{1}|_0 \cap \mathbf{A}
$$
The term $\mathbf{A} ~ {\cup}.{\cap} ~ \mathbb{I}(\mathbf{k}(i))$ selects column $\mathbf{k}(i)$ from $\mathbf{A}$.  The next operation $\cap~v$ selects the entries corresponding to $v$.  A mask of all the columns in these rows is constructed by ${\cup}.{\cap} ~ \mathbs{1}$, whose values are converted to $\mathcal{P}(\mathbb{V})$ with the zero norm $|~|_0$.  Applying the mask with $\cap~\mathbf{A}$ selects the corresponding rows.

\subsection{Deep Neural Networks}

Machine learning has been the foundation of artificial intelligence since its inception \cite{ware1955introduction,clark1955generalization,selfridge1955pattern,dinneen1955programming,newell1955chess,mccarthy2006proposal,minsky1960learning,minsky1961steps}. Standard machine learning applications include speech recognition \cite{selfridge1955pattern}, computer vision \cite{dinneen1955programming}, and even board games \cite{newell1955chess,samuel1959some}.

\begin{figure}[htb]
  	\centering
    	\includegraphics[width=\columnwidth]{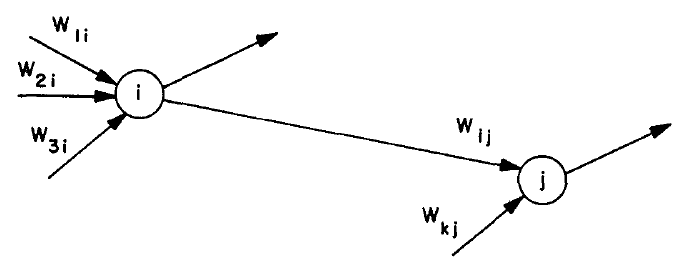}
      	\caption{Typical network elements $i$ and $j$ showing connection weights $w$ (reproduced from  \cite{clark1955generalization})}
      	\label{fig:clark1955fig1}
\end{figure}

\begin{figure}[htb]
  	\centering
    	\includegraphics[width=\columnwidth]{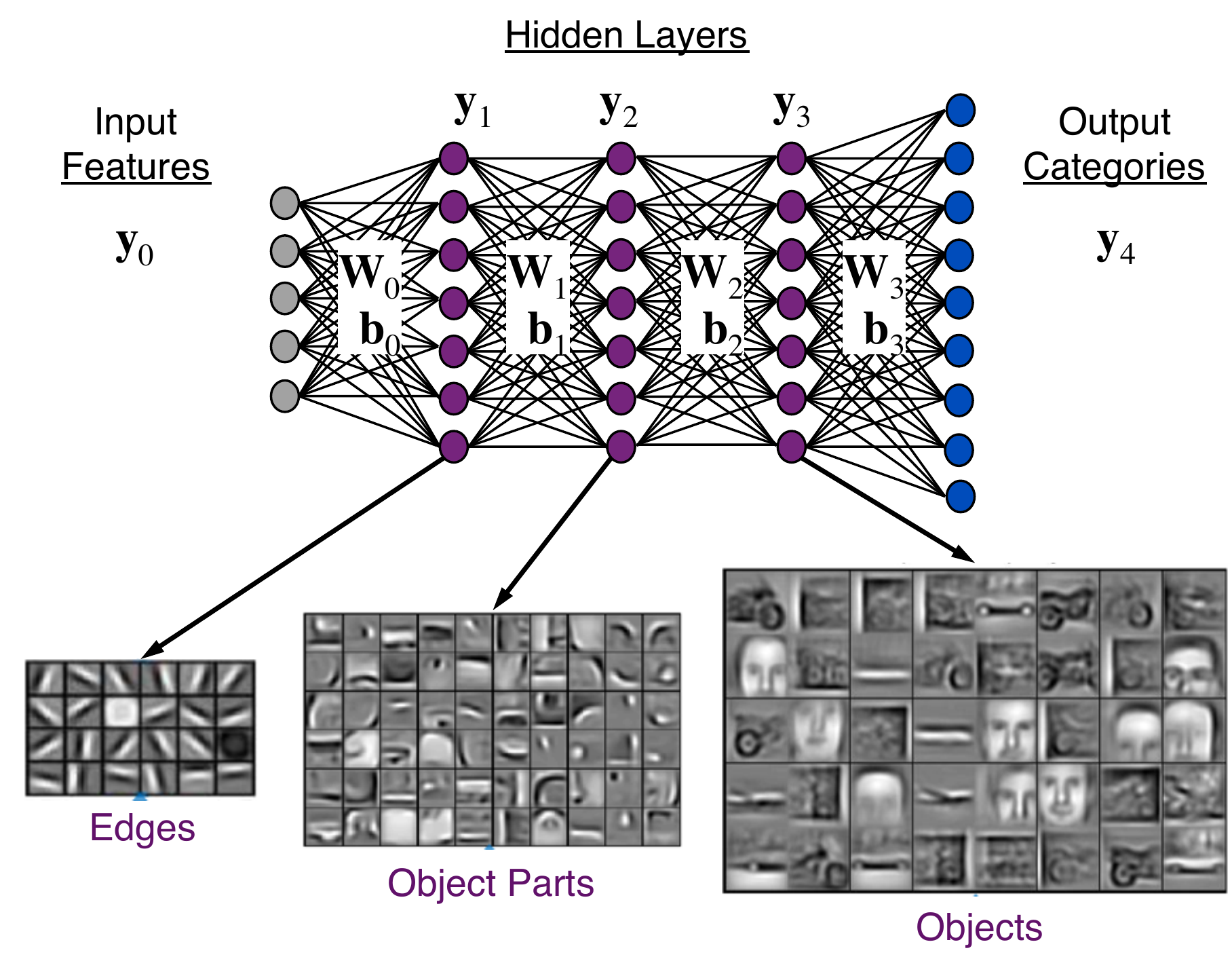}
	\caption{Four layer ($L=4$) deep neural network architecture
	for categorizing images.  The input
	features  ${\bf y}_0$ of an image are passed through a series
	of network layers ${\bf W}_{\ell=0,1,2,3}$, with bias terms
	${\bf b}_{\ell=0,1,2,3}$, that produce scores for categories
	${\bf y}_{L=4}$.  (Figure adapted from \cite{lee2009convolutional})}
      	\label{fig:DNNarchitecture}
\end{figure}

Drawing inspiration from biological neurons to implement machine learning was the topic of the first paper presented at the first machine learning conference in 1955 \cite{ware1955introduction,clark1955generalization} (see Figure~\ref{fig:clark1955fig1}). It was recognized very early on in the field that direct computational training of neural networks was computationally unfeasible with the computers that were available at that time \cite{minsky1960learning}.  The many-fold improvement in neural network computation and theory has made it possible to create neural networks capable of better-than-human performance in a variety of domains \cite{lippmann1987introduction,reynolds2000speaker,krizhevsky2012imagenet,lecun2015deep}. The production of validated data sets \cite{campbell1995testing,lecun1998mnist,deng2009imagenet} and the power of graphic processing units (GPUs) \cite{campbell2002deep,mcgraw2007benchmarking,kerr2008gpu,epstein2012making}
have allowed the effective training of deep neural networks (DNNs) with 100,000s of input features, $N$, and 100s of layers, $L$, that are capable of choosing from among 100,000s categories, $M$ (see Figure~\ref{fig:DNNarchitecture}).

 The primary mathematical operation performed by a DNN network is the inference, or forward propagation, step.  Inference is executed repeatedly during training to determine both the weight matrix ${\bf W}_\ell$ and the bias vectors ${\bf b}_\ell$ of the DNN.  The inference computation shown in Figure~\ref{fig:DNNarchitecture} is given by
$$
  {\bf y}_{\ell + 1} = h({\bf y}_\ell {\bf W}_\ell  + {\bf b}_\ell)
$$
where $h()$ is a nonlinear function applied to each element of the vector.  The Sparse DNN Challenge uses the standard graph community convention whereby ${\bf W}(i,j) \neq 0$  implies a connection between neuron $i$ and neuron $j$.  In this convention ${\bf y}_\ell$ are row vectors and left array multiplication is used to progress through the network.  A commonly used function is the rectified linear unit (ReLU) given by
$$
   h({\bf y}) = \max({\bf y},0)
$$
which sets values less than 0 to 0 and leaves other values unchanged.  When training a DNN, or performing inference on many different inputs, it is usually necessary to compute multiple ${\bf y}_\ell$ vectors at once in a batch that can be denoted as the array ${\bf Y}_\ell$.  In array form, the inference step becomes
$$
  {\bf Y}_{\ell + 1} = h({\bf Y}_\ell {\bf W}_\ell  + {\bf B}_\ell)
$$
where ${\bf B}_\ell$ is a replication of ${\bf b}_\ell$ along columns given by
$$
  \mathbf{B}_\ell = \mathbf{b}_\ell |\mathbf{Y}_\ell \mathbf{1}|_0
$$
and $\mathbf{1}$ is a column array of 1's, and $| ~ |_0$ is the zero norm.

  If $h()$ were a linear function, then the above equation could be solved exactly and the computation could be greatly simplified.  However, current evidence suggests that the non-linearity of $h()$ is required for a DNN to be effective.  Interestingly, the inference computation can be rewritten as a linear function over two different semirings
$$
  {\bf y}_{k + 1} ={\bf y}_k {\bf W}_k \otimes {\bf b}_k \oplus 0
$$
or in array form
$$
  {\bf Y}_{k + 1} ={\bf Y}_k {\bf W}_k  \otimes {\bf B}_k \oplus 0
$$
where the $\oplus = \max$ and $\otimes = +$.  Thus, ${\bf y}_k {\bf W}_k$ and ${\bf Y}_k {\bf W}_k$ are computed over the standard arithmetic ${+}.{\times}$ semiring 
$$
  S_1 = (\mathbb{R},+,\times,0,1)
$$
while the $\oplus$ and $\otimes$  operation are performed over the ${\max}.{+}$ semiring
$$
  S_2 = (\{ \text{-}\infty \cup \mathbb{R} \},\max,+,\text{-}\infty,0)
$$
Thus, the ReLU DNN can be written as a linear system that oscillates over two semirings $S_1$ and $S_2$.  $S_1$ is the most widely used of semirings and performs standard correlation between vectors.  $S_2$ is also a commonly used semiring for selecting optimal paths in graphs.  Thus, the inference step of a ReLU DNN can be viewed as combining correlations of inputs to choose optimal paths through the neural network.  This DNN semiring pair is is more complex than what is described in by the semilink concept and may require extending the semilink concept to encompass DNNs.

\section{Conclusions and Future Work}

The unstructured data of digital hyperspace can be elegantly represented, traversed, and transformed via the mathematics of hypergraphs, hypersparse matrices, and associative array algebra.  Within this context this paper has explored a new mathematical concept, the semilink, that combines pairs of semirings to provide the essential operations for graph analytics, database operations, and machine learning.  The formal mathematical specification of GraphBLAS includes monoid, semiring, and closure under element-wise addition, element-wise multiplication, and array multiplication and naturally supports linked semirings.

The specification was written from an associative array algebra perspective with intentionally minimal constraints on the internal implementation of the opaque GrB\_Matrix data structure.  This has allowed the GraphBLAS (in its SuiteSparse implementation) to support a myriad of different data structures: sparse, hypersparse, bitmap, and full.  It uses each of them when appropriate, and switches between them automatically, with little or no involvement from the user application.  In the future, this will enable distributed-memory and GPU accelerations as well.  This flexibility has enabled the GraphBLAS standard to support hypergraphs, hypersparse matrices, and the mathematics required for semilinks, and seamlessly performs graph, network, and matrix operations.  With the addition of key based indices (such as pointers to strings) and semilinks, GraphBLAS can become a richer associative array algebra and be a plug-in replacement for spreadsheets, database tables, and data centric operating systems \cite{kepner2018tabularosa,cafarella2020dbos}, enhancing the navigation of unstructured data found in digital hyperspace.

From an applied mathematical perspective, the more complex pairing of operations in the DNN context is worth additional exploring.  Likewise, in the context of abstract algebra, \cite{wilding2015linear} considers semirings in which the multiplicative identity can be local, so that in any small part of the structure there is a multiplicative identity as far as that part of the structure is concerned.  It would be worth exploring this concept in the context of infinite key spaces where identity matrices are a challenge.

\section*{Acknowledgments}

The authors wish to acknowledge the following individuals for their contributions and support: Bob Bond, Alan Edelman, Jeff Gottschalk, Charles Leiserson, Mimi McClure, Steve Rejto, Daniela Rus, Allan Vanterpool, Marc Zissman, and the MIT SuperCloud team: Bill Arcand, Bill Bergeron, David Bestor, Chansup Byun, Michael Houle, Matthew Hubbell, Michael Jones, Anna Klein, Peter Michaleas, Julie Mullen, Andrew Prout, Antonio Rosa, Albert Reuther, Charles Yee.

\bibliographystyle{ieeetr}
\bibliography{NavigatingDigitalHyperspace}

\begin{thebibliography}{10}

\bibitem{Cisco2018-2023}
``{\it Cisco Visual Networking Index: Forecast and Trends, 2018–2023}.''
  https://www.cisco.com/c/en/us/solutions/collateral/executive-perspectives/annual-internet-report/white-paper-c11-741490.html.

\bibitem{sawadogo2020data}
P.~Sawadogo and J.~Darmont, ``On data lake architectures and metadata
  management,'' {\em Journal of Intelligent Information Systems}, pp.~1--24,
  2020.

\bibitem{ghoshal2009random}
G.~Ghoshal, V.~Zlati{\'c}, G.~Caldarelli, and M.~E. Newman, ``Random
  hypergraphs and their applications,'' {\em Physical Review E}, vol.~79,
  no.~6, p.~066118, 2009.

\bibitem{mordeson2012fuzzy}
J.~N. Mordeson and P.~S. Nair, {\em Fuzzy graphs and fuzzy hypergraphs},
  vol.~46.
\newblock Physica, 2012.

\bibitem{shun2020practical}
J.~Shun, ``Practical parallel hypergraph algorithms,'' in {\em Proceedings of
  the 25th ACM SIGPLAN Symposium on Principles and Practice of Parallel
  Programming}, pp.~232--249, 2020.

\bibitem{buluc2008representation}
A.~Buluc and J.~R. Gilbert, ``On the representation and multiplication of
  hypersparse matrices,'' in {\em 2008 IEEE International Symposium on Parallel
  and Distributed Processing}, pp.~1--11, IEEE, 2008.

\bibitem{kepner2011graph}
J.~Kepner and J.~Gilbert, {\em Graph algorithms in the language of linear
  algebra}.
\newblock SIAM, 2011.

\bibitem{kepner202075}
J.~Kepner, T.~Davis, C.~Byun, W.~Arcand, D.~Bestor, W.~Bergeron, V.~Gadepally,
  M.~Hubbell, M.~Houle, M.~Jones, A.~Klein, P.~Michaleas, L.~Milechin,
  J.~Mullen, A.~Prout, A.~Rosa, S.~Samsi, C.~Yee, and A.~Reuther,
  ``75,000,000,000 streaming inserts/second using hierarchical hypersparse
  graphblas matrices,'' {\em IPDPSW GrAPL}, 2020.

\bibitem{Kepner2011p-ch1}
J.~V. Kepner, ``Multidimensional associative array database,'' Jan.~14 2014.
\newblock US Patent 8,631,031.

\bibitem{KepnerGadepally2014}
J.~Kepner and V.~Gadepally, ``Adjacency matrices, incidence matrices, database
  schemas, and associative arrays,'' {\em IPDPS Graph Algorithms Building
  Blocks}, 2014.

\bibitem{kepner2016associative}
J.~{Kepner}, V.~{Gadepally}, D.~{Hutchison}, H.~{Jananthan}, T.~{Mattson},
  S.~{Samsi}, and A.~{Reuther}, ``Associative array model of sql, nosql, and
  newsql databases,'' in {\em 2016 IEEE High Performance Extreme Computing
  Conference (HPEC)}, pp.~1--9, 2016.

\bibitem{kepner2018mathematics}
J.~Kepner and H.~Jananthan, {\em Mathematics of Big Data: Spreadsheets,
  databases, matrices, and graphs}.
\newblock MIT Press, 2018.

\bibitem{Kepner2016graphblas}
J.~Kepner, P.~Aaltonen, D.~Bader, A.~Bulu{\c{c}}, F.~Franchetti, J.~Gilbert,
  D.~Hutchison, M.~Kumar, A.~Lumsdaine, H.~Meyerhenke, S.~McMillan, J.~Moreira,
  J.~Owens, C.~Yang, M.~Zalewski, and T.~Mattson, ``Mathematical foundations of
  the {GraphBLAS},'' in {\em High Performance Extreme Computing Conference
  (HPEC)}, IEEE, 2016.

\bibitem{bulucc2017design}
A.~Bulu{\c{c}}, T.~Mattson, S.~McMillan, J.~Moreira, and C.~Yang, ``Design of
  the graphblas api for c,'' in {\em 2017 IEEE International Parallel and
  Distributed Processing Symposium Workshops (IPDPSW)}, pp.~643--652, IEEE,
  2017.

\bibitem{kumar2018graphblas}
M.~Kumar, J.~E. Moreira, and P.~Pattnaik, ``Graphblas: handling performance
  concerns in large graph analytics,'' in {\em Proceedings of the 15th ACM
  International Conference on Computing Frontiers}, pp.~260--267, 2018.

\bibitem{mattson2019lagraph}
T.~Mattson, T.~A. Davis, M.~Kumar, A.~Buluc, S.~McMillan, J.~Moreira, and
  C.~Yang, ``Lagraph: A community effort to collect graph algorithms built on
  top of the graphblas,'' in {\em 2019 IEEE International Parallel and
  Distributed Processing Symposium Workshops (IPDPSW)}, pp.~276--284, IEEE,
  2019.

\bibitem{davis2018graph}
T.~A. Davis, ``Graph algorithms via suitesparse: Graphblas: triangle counting
  and k-truss,'' in {\em 2018 IEEE High Performance extreme Computing
  Conference (HPEC)}, pp.~1--6, IEEE, 2018.

\bibitem{chamberlin2018pygb}
J.~Chamberlin, M.~Zalewski, S.~McMillan, and A.~Lumsdaine, ``Pygb: Graphblas
  dsl in python with dynamic compilation into efficient c++,'' in {\em 2018
  IEEE International Parallel and Distributed Processing Symposium Workshops
  (IPDPSW)}, pp.~310--319, IEEE, 2018.

\bibitem{moreira2018implementing}
J.~E. Moreira, M.~Kumar, and W.~P. Horn, ``Implementing the graphblas c api,''
  in {\em 2018 IEEE International Parallel and Distributed Processing Symposium
  Workshops (IPDPSW)}, pp.~298--309, IEEE, 2018.

\bibitem{davis2019algorithm}
T.~A. Davis, ``Algorithm 1000: Suitesparse: Graphblas: Graph algorithms in the
  language of sparse linear algebra,'' {\em ACM Transactions on Mathematical
  Software (TOMS)}, vol.~45, no.~4, pp.~1--25, 2019.

\bibitem{cailliau2019redisgraph}
P.~Cailliau, T.~Davis, V.~Gadepally, J.~Kepner, R.~Lipman, J.~Lovitz, and
  K.~Ouaknine, ``Redisgraph graphblas enabled graph database,'' in {\em 2019
  IEEE International Parallel and Distributed Processing Symposium Workshops
  (IPDPSW)}, pp.~285--286, IEEE, 2019.

\bibitem{BulucGilbert2011}
A.~Bulu{\c{c}} and J.~R. Gilbert, ``The combinatorial blas: Design,
  implementation, and applications,'' {\em The International Journal of High
  Performance Computing Applications}, vol.~25, no.~4, pp.~496--509, 2011.

\bibitem{kepner2011massive}
J.~Kepner, W.~Arcand, W.~Bergeron, C.~Byun, M.~Hubbell, B.~Landon, A.~McCabe,
  P.~Michaleas, A.~Prout, T.~Rosa, {\em et~al.}, ``Massive database analysis on
  the cloud with d4m,'' {\em HPEC, Sep}, pp.~21--22, 2011.

\bibitem{Kepner2012-ch1}
J.~Kepner, W.~Arcand, W.~Bergeron, N.~Bliss, R.~Bond, C.~Byun, G.~Condon,
  K.~Gregson, M.~Hubbell, J.~Kurz, A.~McCabe, P.~Michaleas, A.~Prout,
  A.~Reuther, A.~Rosa, and C.~Yee, ``Dynamic distributed dimensional data model
  {(D4M)} database and computation system,'' in {\em Acoustics, Speech and
  Signal Processing (ICASSP), 2012 IEEE International Conference on},
  pp.~5349--5352, IEEE, 2012.

\bibitem{chen2016julia}
A.~Chen, A.~Edelman, J.~Kepner, V.~Gadepally, and D.~Hutchison, ``Julia
  implementation of the dynamic distributed dimensional data model,'' in {\em
  2016 IEEE High Performance Extreme Computing Conference (HPEC)}, pp.~1--7,
  IEEE, 2016.

\bibitem{milechin2017d4m}
L.~Milechin, V.~Gadepally, S.~Samsi, J.~Kepner, A.~Chen, and D.~Hutchison,
  ``D4m 3.0: Extended database and language capabilities,'' in {\em 2017 IEEE
  High Performance Extreme Computing Conference (HPEC)}, pp.~1--6, IEEE, 2017.

\bibitem{milechin2018database}
L.~Milechin, V.~Gadepally, and J.~Kepner, ``Database operations in d4m.jl,'' in
  {\em 2018 IEEE High Performance extreme Computing Conference (HPEC)},
  pp.~1--5, IEEE, 2018.

\bibitem{Kepner2013}
J.~Kepner, C.~Anderson, W.~Arcand, D.~Bestor, B.~Bergeron, C.~Byun, M.~Hubbell,
  P.~Michaleas, J.~Mullen, D.~O'Gwynn, A.~Prout, A.~Reuther, A.~Rosa, and
  C.~Yee, ``{D4M} 2.0 schema: A general purpose high performance schema for the
  {Accumulo} database,'' in {\em High Performance Extreme Computing Conference
  (HPEC)}, IEEE, 2013.

\bibitem{Kepner2014a}
J.~Kepner, W.~Arcand, D.~Bestor, B.~Bergeron, C.~Byun, V.~Gadepally,
  M.~Hubbell, P.~Michaleas, J.~Mullen, A.~Prout, {\em et~al.}, ``Achieving
  100,000,000 database inserts per second using {Accumulo} and {D4M},'' in {\em
  High Performance Extreme Computing Conference (HPEC)}, IEEE, 2014.

\bibitem{GadepallyEtAl2015}
V.~Gadepally, J.~Kepner, W.~Arcand, D.~Bestor, B.~Bergeron, C.~Byun,
  L.~Edwards, M.~Hubbell, P.~Michaleas, J.~Mullen, {\em et~al.}, ``{D4M:}
  bringing associative arrays to database engines,'' in {\em High Performance
  Extreme Computing Conference (HPEC)}, IEEE, 2015.

\bibitem{Hutchison2015}
D.~Hutchison, J.~Kepner, V.~Gadepally, and A.~Fuchs, ``Graphulo implementation
  of server-side sparse matrix multiply in the {Accumulo} database,'' in {\em
  High Performance Extreme Computing Conference (HPEC)}, IEEE, 2015.

\bibitem{samsi2016benchmarking}
S.~Samsi, L.~Brattain, W.~Arcand, D.~Bestor, B.~Bergeron, C.~Byun,
  V.~Gadepally, M.~Hubbell, M.~Jones, A.~Klein, {\em et~al.}, ``Benchmarking
  scidb data import on hpc systems,'' in {\em 2016 IEEE High Performance
  Extreme Computing Conference (HPEC)}, pp.~1--5, IEEE, 2016.

\bibitem{aznaveh2020parallel}
M.~Aznaveh, J.~Chen, T.~A. Davis, B.~Hegyi, S.~P. Kolodziej, T.~G. Mattson, and
  G.~Szarnyas, ``Parallel graphblas with openmp,'' in {\em 2020 Proceedings of
  the SIAM Workshop on Combinatorial Scientific Computing}, pp.~138--148, SIAM,
  2020.

\bibitem{wang2019accelerating}
X.~Wang, Z.~Lin, C.~Yang, and J.~D. Owens, ``Accelerating dnn inference with
  graphblas and the gpu,'' in {\em 2019 IEEE High Performance Extreme Computing
  Conference (HPEC)}, pp.~1--6, IEEE, 2019.

\bibitem{song2010}
W.~S. Song, J.~Kepner, H.~T. Nguyen, J.~I. Kramer, V.~Gleyzer, J.~R. Mann,
  A.~H. Horst, L.~L. Retherford, R.~A. Bond, N.~T. Bliss, {\em et~al.}, ``3-d
  graph processor,'' in {\em Workshop on High Performance Embedded Workshop
  (HPEC)}, MIT Lincoln Laboratory, 2010.

\bibitem{song2014processor}
W.~S. Song, ``Processor for large graph algorithm computations and matrix
  operations,'' June~10 2014.
\newblock US Patent 8,751,556.

\bibitem{song2016novel}
W.~S. Song, V.~Gleyzer, A.~Lomakin, and J.~Kepner, ``Novel graph processor
  architecture, prototype system, and results,'' in {\em High Performance
  Extreme Computing Conference (HPEC)}, IEEE, 2016.

\bibitem{jia2019dissecting}
Z.~Jia, B.~Tillman, M.~Maggioni, and D.~P. Scarpazza, ``Dissecting the
  graphcore ipu architecture via microbenchmarking,'' {\em arXiv preprint
  arXiv:1912.03413}, 2019.

\bibitem{james2020ispd}
M.~James, M.~Tom, P.~Groeneveld, and V.~Kibardin, ``Ispd 2020 physical mapping
  of neural networks on a wafer-scale deep learning accelerator,'' in {\em
  Proceedings of the 2020 International Symposium on Physical Design},
  pp.~145--149, 2020.

\bibitem{kepner2017enabling}
J.~Kepner, M.~Kumar, J.~Moreira, P.~Pattnaik, M.~Serrano, and H.~Tufo,
  ``Enabling massive deep neural networks with the graphblas,'' in {\em High
  Performance Extreme Computing Conference (HPEC)}, IEEE, 2017.

\bibitem{kumar2018ibm}
M.~Kumar, W.~Horn, J.~Kepner, J.~Moreira, and P.~Pattnaik, ``Ibm power9 and
  cognitive computing,'' {\em IBM Journal of Research and Development}, 2018.

\bibitem{davis2019write}
T.~A. Davis, M.~Aznaveh, and S.~Kolodziej, ``Write quick, run fast: Sparse deep
  neural network in 20 minutes of development time via suitesparse:
  Graphblas,'' in {\em 2019 IEEE High Performance extreme Computing Conference
  (HPEC)}, pp.~1--6, IEEE, 2019.

\bibitem{codd1970relational}
E.~F. Codd, ``A relational model of data for large shared data banks,'' {\em
  Communications of the ACM}, vol.~13, no.~6, pp.~377--387, 1970.

\bibitem{maier1983theory}
D.~Maier, {\em The theory of relational databases}, vol.~11.
\newblock Computer science press Rockville, 1983.

\bibitem{abiteboul1995foundations}
S.~Abiteboul, R.~Hull, and V.~Vianu, {\em Foundations of databases}, vol.~8.
\newblock Addison-Wesley Reading, 1995.

\bibitem{klemperer2010product}
P.~Klemperer, ``The product-mix auction: A new auction design for
  differentiated goods,'' {\em Journal of the European Economic Association},
  vol.~8, no.~2-3, pp.~526--536, 2010.

\bibitem{baldwin2016understanding}
E.~Baldwin and P.~Klemperer, ``Understanding preferences:'demand types', and
  the existence of equilibrium with indivisibilities,'' {\em SSRN}, 2016.

\bibitem{glazek2002guide}
K.~Glazek, {\em A guide to the literature on semirings and their applications
  in mathematics and information sciences: with complete bibliography}.
\newblock Springer Science \& Business Media, 2002.

\bibitem{blount2003structure}
K.~Blount and C.~Tsinakis, ``The structure of residuated lattices,'' {\em
  International Journal of Algebra and Computation}, vol.~13, no.~04,
  pp.~437--461, 2003.

\bibitem{aguiar2000pre}
M.~Aguiar, ``Pre-poisson algebras,'' {\em Letters in Mathematical Physics},
  vol.~54, no.~4, pp.~263--277, 2000.

\bibitem{kuhlmann2000ordered}
S.~Kuhlmann, {\em Ordered exponential fields}, vol.~12.
\newblock American Mathematical Soc., 2000.

\bibitem{smith2006introduction}
J.~D. Smith, {\em An introduction to quasigroups and their representations}.
\newblock CRC Press, 2006.

\bibitem{Stonebraker1976}
M.~Stonebraker, G.~Held, E.~Wong, and P.~Kreps, ``The design and implementation
  of {INGRES},'' {\em ACM Transactions on Database Systems (TODS)}, vol.~1,
  no.~3, pp.~189--222, 1976.

\bibitem{date1989guide}
C.~J. Date and H.~Darwen, {\em A guide to the SQL Standard: a user's guide to
  the standard relational language SQL}.
\newblock Addison-Wesley, 1989.

\bibitem{elmasri2010fundamentals}
R.~Elmasri and S.~Navathe, {\em Fundamentals of database systems}.
\newblock Addison-Wesley Publishing Company, 2010.

\bibitem{jananthan2017polystore}
H.~Jananthan, Z.~Zhou, V.~Gadepally, D.~Hutchison, S.~Kim, and J.~Kepner,
  ``Polystore mathematics of relational algebra,'' in {\em Big Data Workshop on
  Methods to Manage Heterogeneous Big Data and Polystore Databases}, IEEE,
  2017.

\bibitem{DeCandia2007}
G.~DeCandia, D.~Hastorun, M.~Jampani, G.~Kakulapati, A.~Lakshman, A.~Pilchin,
  S.~Sivasubramanian, P.~Vosshall, and W.~Vogels, ``Dynamo: amazon's highly
  available key-value store,'' {\em ACM SIGOPS operating systems review},
  vol.~41, no.~6, pp.~205--220, 2007.

\bibitem{LakshmanMalik2010}
A.~Lakshman and P.~Malik, ``Cassandra: a decentralized structured storage
  system,'' {\em ACM SIGOPS Operating Systems Review}, vol.~44, no.~2,
  pp.~35--40, 2010.

\bibitem{George2011}
L.~George, {\em HBase: The Definitive Guide: Random Access to Your Planet-Size
  Data}.
\newblock " O'Reilly Media, Inc.", 2011.

\bibitem{7322476}
J.~{Kepner}, W.~{Arcand}, D.~{Bestor}, B.~{Bergeron}, C.~{Byun}, L.~{Edwards},
  V.~{Gadepally}, M.~{Hubbell}, P.~{Michaleas}, J.~{Mullen}, A.~{Prout},
  A.~{Rosa}, C.~{Yee}, and A.~{Reuther}, ``Lustre, hadoop, accumulo,'' in {\em
  2015 IEEE High Performance Extreme Computing Conference (HPEC)}, pp.~1--5,
  Sep. 2015.

\bibitem{Wall2015}
A.~Cordova, B.~Rinaldi, and M.~Wall, {\em Accumulo: Application Development,
  Table Design, and Best Practices}.
\newblock " O'Reilly Media, Inc.", 2015.

\bibitem{Stonebraker2005}
M.~Stonebraker, D.~J. Abadi, A.~Batkin, X.~Chen, M.~Cherniack, M.~Ferreira,
  E.~Lau, A.~Lin, S.~Madden, E.~O'Neil, {\em et~al.}, ``{C-Store:} a
  column-oriented {DBMS},'' in {\em Proceedings of the 31st international
  conference on Very large data bases}, pp.~553--564, VLDB Endowment, 2005.

\bibitem{Kallman2008}
R.~Kallman, H.~Kimura, J.~Natkins, A.~Pavlo, A.~Rasin, S.~Zdonik, E.~P. Jones,
  S.~Madden, M.~Stonebraker, Y.~Zhang, {\em et~al.}, ``H-store: a
  high-performance, distributed main memory transaction processing system,''
  {\em Proceedings of the VLDB Endowment}, vol.~1, no.~2, pp.~1496--1499, 2008.

\bibitem{Balazinska2009}
P.~Cudr{\'e}-Mauroux, H.~Kimura, K.-T. Lim, J.~Rogers, R.~Simakov, E.~Soroush,
  P.~Velikhov, D.~L. Wang, M.~Balazinska, J.~Becla, J.~Becla, D.~DeWitt,
  B.~Heath, D.~Maier, S.~Madden, J.~Patel, M.~Stonebraker, and S.~Zdonik, ``A
  demonstration of {SciDB:} a science-oriented {DBMS},'' {\em Proceedings of
  the VLDB Endowment}, vol.~2, no.~2, pp.~1534--1537, 2009.

\bibitem{StonebrakerWeisberg2013}
M.~Stonebraker and A.~Weisberg, ``The {VoltDB} main memory {DBMS},'' {\em IEEE
  Data Engineering Bulletin}, vol.~36, no.~2, pp.~21--27, 2013.

\bibitem{gadepally2015graphulo}
V.~Gadepally, J.~Bolewski, D.~Hook, D.~Hutchison, B.~Miller, and J.~Kepner,
  ``Graphulo: Linear algebra graph kernels for nosql databases,'' in {\em
  Parallel and Distributed Processing Symposium Workshop (IPDPSW), 2015 IEEE
  International}, pp.~822--830, IEEE, 2015.

\bibitem{booth1976distributed}
G.~M. Booth, ``Distributed information systems,'' in {\em Proceedings of the
  June 7-10, 1976, national computer conference and exposition}, pp.~789--794,
  ACM, 1976.

\bibitem{shaw1980relational}
D.~E. Shaw, ``A relational database machine architecture,'' in {\em ACM SIGIR
  Forum}, vol.~15 \#2, pp.~84--95, ACM, 1980.

\bibitem{stonebraker1986case}
M.~Stonebraker, ``The case for shared nothing,'' {\em IEEE Database Eng.
  Bull.}, vol.~9, no.~1, pp.~4--9, 1986.

\bibitem{barroso2003web}
L.~A. Barroso, J.~Dean, and U.~Holzle, ``Web search for a planet: The google
  cluster architecture,'' {\em IEEE micro}, vol.~23, no.~2, pp.~22--28, 2003.

\bibitem{curino2010schism}
C.~Curino, E.~Jones, Y.~Zhang, and S.~Madden, ``Schism: a workload-driven
  approach to database replication and partitioning,'' {\em Proceedings of the
  VLDB Endowment}, vol.~3, no.~1-2, pp.~48--57, 2010.

\bibitem{pavlo2012skew}
A.~Pavlo, C.~Curino, and S.~Zdonik, ``Skew-aware automatic database
  partitioning in shared-nothing, parallel oltp systems,'' in {\em Proceedings
  of the 2012 ACM SIGMOD International Conference on Management of Data},
  pp.~61--72, ACM, 2012.

\bibitem{corbett2013spanner}
J.~C. Corbett, J.~Dean, M.~Epstein, A.~Fikes, C.~Frost, J.~J. Furman,
  S.~Ghemawat, A.~Gubarev, C.~Heiser, P.~Hochschild, {\em et~al.}, ``Spanner:
  Google's globally distributed database,'' {\em ACM Transactions on Computer
  Systems (TOCS)}, vol.~31, no.~3, p.~8, 2013.

\bibitem{ware1955introduction}
W.~H. Ware, ``Introduction to session on learning machines,'' in {\em
  Proceedings of the March 1-3, 1955, western joint computer conference},
  pp.~85--85, ACM, 1955.

\bibitem{clark1955generalization}
W.~A. Clark and B.~G. Farley, ``Generalization of pattern recognition in a
  self-organizing system,'' in {\em Proceedings of the March 1-3, 1955, western
  joint computer conference}, pp.~86--91, ACM, 1955.

\bibitem{selfridge1955pattern}
O.~G. Selfridge, ``Pattern recognition and modern computers,'' in {\em
  Proceedings of the March 1-3, 1955, western joint computer conference},
  pp.~91--93, ACM, 1955.

\bibitem{dinneen1955programming}
G.~Dinneen, ``Programming pattern recognition,'' in {\em Proceedings of the
  March 1-3, 1955, western joint computer conference}, pp.~94--100, ACM, 1955.

\bibitem{newell1955chess}
A.~Newell, ``The chess machine: an example of dealing with a complex task by
  adaptation,'' in {\em Proceedings of the March 1-3, 1955, western joint
  computer conference}, pp.~101--108, ACM, 1955.

\bibitem{mccarthy2006proposal}
J.~McCarthy, M.~L. Minsky, N.~Rochester, and C.~E. Shannon, ``A proposal for
  the dartmouth summer research project on artificial intelligence, august 31,
  1955,'' {\em AI magazine}, vol.~27, no.~4, p.~12, 2006.

\bibitem{minsky1960learning}
M.~Minsky and O.~G. Selfridge, ``Learning in random nets,'' in {\em Information
  theory : papers read at a symposium on information theory held at the Royal
  Institution, London, August 29th to September 2nd}, pp.~335--347,
  Butterworths, London, 1960.

\bibitem{minsky1961steps}
M.~Minsky, ``Steps toward artificial intelligence,'' {\em Proceedings of the
  IRE}, vol.~49, no.~1, pp.~8--30, 1961.

\bibitem{samuel1959some}
A.~L. Samuel, ``Some studies in machine learning using the game of checkers,''
  {\em IBM Journal of research and development}, vol.~3, no.~3, pp.~210--229,
  1959.

\bibitem{lippmann1987introduction}
R.~Lippmann, ``An introduction to computing with neural nets,'' {\em IEEE Assp
  magazine}, vol.~4, no.~2, pp.~4--22, 1987.

\bibitem{reynolds2000speaker}
D.~A. Reynolds, T.~F. Quatieri, and R.~B. Dunn, ``Speaker verification using
  adapted gaussian mixture models,'' {\em Digital signal processing}, vol.~10,
  no.~1-3, pp.~19--41, 2000.

\bibitem{krizhevsky2012imagenet}
A.~Krizhevsky, I.~Sutskever, and G.~E. Hinton, ``Imagenet classification with
  deep convolutional neural networks,'' in {\em Advances in neural information
  processing systems}, pp.~1097--1105, 2012.

\bibitem{lecun2015deep}
Y.~LeCun, Y.~Bengio, and G.~Hinton, ``Deep learning,'' {\em Nature}, vol.~521,
  no.~7553, pp.~436--444, 2015.

\bibitem{campbell1995testing}
J.~P. Campbell, ``Testing with the yoho cd-rom voice verification corpus,'' in
  {\em Acoustics, Speech, and Signal Processing, 1995. ICASSP-95., 1995
  International Conference on}, vol.~1, pp.~341--344, IEEE, 1995.

\bibitem{lecun1998mnist}
Y.~LeCun, C.~Cortes, and C.~J. Burges, ``The mnist database of handwritten
  digits,'' 1998.

\bibitem{deng2009imagenet}
J.~Deng, W.~Dong, R.~Socher, L.-J. Li, K.~Li, and L.~Fei-Fei, ``Imagenet: A
  large-scale hierarchical image database,'' in {\em Computer Vision and
  Pattern Recognition, 2009. CVPR 2009. IEEE Conference on}, pp.~248--255,
  IEEE, 2009.

\bibitem{campbell2002deep}
M.~Campbell, A.~J. Hoane, and F.-h. Hsu, ``Deep blue,'' {\em Artificial
  intelligence}, vol.~134, no.~1-2, pp.~57--83, 2002.

\bibitem{mcgraw2007benchmarking}
M.~P. McGraw-Herdeg, D.~P. Enright, and B.~S. Michel, ``Benchmarking the nvidia
  8800gtx with the cuda development platform,'' {\em HPEC 2007 Proceedings},
  2007.

\bibitem{kerr2008gpu}
A.~Kerr, D.~Campbell, and M.~Richards, ``Gpu performance assessment with the
  hpec challenge,'' in {\em HPEC Workshop 2008}, 2008.

\bibitem{epstein2012making}
E.~A. Epstein, M.~I. Schor, B.~Iyer, A.~Lally, E.~W. Brown, and J.~Cwiklik,
  ``Making watson fast,'' {\em IBM Journal of Research and Development},
  vol.~56, no.~3.4, pp.~15--1, 2012.

\bibitem{lee2009convolutional}
H.~Lee, R.~Grosse, R.~Ranganath, and A.~Y. Ng, ``Convolutional deep belief
  networks for scalable unsupervised learning of hierarchical
  representations,'' in {\em Proceedings of the 26th annual international
  conference on machine learning}, pp.~609--616, ACM, 2009.

\bibitem{kepner2018tabularosa}
J.~Kepner, R.~Brightwell, A.~Edelman, V.~Gadepally, H.~Jananthan, M.~Jones,
  S.~Madden, P.~Michaleas, H.~Okhravi, K.~Pedretti, {\em et~al.}, ``Tabularosa:
  Tabular operating system architecture for massively parallel heterogeneous
  compute engines,'' in {\em 2018 IEEE High Performance extreme Computing
  Conference (HPEC)}, IEEE, 2018.

\bibitem{cafarella2020dbos}
M.~Cafarella, D.~DeWitt, V.~Gadepally, J.~Kepner, C.~Kozyrakis, T.~Kraska,
  M.~Stonebraker, and M.~Zaharia, ``Dbos: A proposal for a data-centric
  operating system,'' {\em arXiv preprint arXiv:2007.11112}, 2020.

\bibitem{wilding2015linear}
D.~Wilding, {\em Linear algebra over semirings}.
\newblock The University of Manchester (United Kingdom), 2015.

\end{thebibliography}

\appendices
\setcounter{equation}{0}
\renewcommand{\theequation}{\thesection\arabic{equation}}

\end{document}